\documentclass[tightenlines,aps,pra,showpacs,twocolumn]{revtex4}
\usepackage{amssymb}
\usepackage{amsmath}
\usepackage{amsfonts}
\usepackage{mathrsfs, mathbbol}
\usepackage{graphicx}
 \usepackage{verbatim}
\newcommand{\ket}[1]{| #1 \: \rangle}
\newcommand{\bra}[1]{\langle \: #1 |}
\newcommand{\out}[2]{{\ket{#1}\bra{#2}}}
\newcommand{\inner}[2]{{\langle \: #1 |  #2 \: \rangle}}

\newcommand{\bea}{\begin{eqnarray}}
\newcommand{\eea}{\end{eqnarray}}
\newcommand{\Tr}[1]{{\rm Tr} \left( #1 \right)}
\newcommand{\pTr}[2]{{\rm Tr}_{#1} \left( #2 \right)}
\newcommand{\lr}[1]{ \left( #1 \right)}
\newcommand{\com}[2]{ \left[ #1,#2 \right]}
\newcommand{\acom}[2]{ \{ #1,#2 \}}
\newcommand{\lra}[1]{ \left| #1 \right|}
\newcommand{\lrb}[1]{ \left[ #1 \right]}

\def \r{\rho}
\def \k{\kappa}
\def \a{\alpha}
\def \b{\beta}

\def \s{\sigma}
\def \o{\omega}
\def \O{\Omega}
\def \ah{\hat{a}}
\def \ahd{\hat{a}^{\dagger}}

\def \nn{\nonumber}
\def \nnl{\nonumber\\}

\begin{document}

\title{Tripartite entanglement dynamics in a system of strongly driven qubits }

\author{Marcin Dukalski and Ya. M. Blanter}
\affiliation{Kavli Institute of Nanoscience, Delft University of Technology, Lorentzweg 1, 2628 CJ Delft, The Netherlands}
\begin{abstract}
We study the dynamics of tripartite entanglement in a system of two strongly driven qubits individually coupled to a dissipative cavity. We aim at explanation of the previously noted entanglement revival between two qubits in this system. We show that the periods of entanglement loss correspond to the strong tripartite entanglement between the qubits and the cavity and the recovery has to do with an inverse process. We demonstrate that the overall process of qubit-qubit entanglement loss is due to the second order coupling to the external continuum which explains the $e^{-g^2 t/2+g^2 \k t^3/6+\cdots}$ for of the entanglement loss reported previously.
\end{abstract}

\maketitle
\section{Introduction}
Entanglement is one of the key aspects distinguishing quantum from classical physics. Its fragileness, due to inevitable coupling of a quantum system, such as  qubits or photons, to a classical environment, however sets limits to its applicability to quantum information and quantum communication technologies. It is therefore very important to understand what is entanglement most vulnerable to and what processes can avert or undo entanglement loss.

Qubits are the fundamental building blocks of quantum information science, and the last two decades marked great developments both theoretically as well as experimentally  \cite{NazarovBlanterbook,  Nielsen}. After many successes in that field, the current research frontier is the qubit-qubit entanglement, which requires either a direct coupling between qubits or an indirect one through an auxiliary system, for example a resonator \cite{JC,Dutra}.  As a result of that the qubits can potentially entangle  to, or disentangle  from, each other depending on the system  design, its  parameters, interaction time, or even the type of environments  \cite{J.H.Eberly04272007, PhysRevB.66.193306, PhysRevLett.93.140404,
PhysRevLett.99.180504,PhysRevA.77.033839, Bina2008, ESB3,ESB1,Dukalski2010,Dukalski2011, M.P.Almeida04272007,fanchini,maniscalco,gordon}.

In this article we focus on the process of entanglement revival in a system of two qubits driven by a strong external, classical ac field and simultaneously coupled a quantum resonator which indirectly couples the qubits. We previously discussed this phenomenon in Ref. \cite{Dukalski2010,Dukalski2011} and demonstrated that entanglement does not only need to decay as the system evolves in time, but the system can also periodically regain some of its initial entanglement. Previously, it was shown for this system that the disentanglement between the qubits may be a consequence of the cavity dissipation  \cite{PhysRevA.77.033839, Bina2008}. In this manuscript, we demonstrate that this is is not the only mechanism leading to entanglement loss. Specifically,
we show that the mere presence of a qubit-cavity coupling results in disentanglement in the subspace spanned by the qubits and that the further coupling of the cavity to the electromagnetic continuum   leads to an overall tripartite entanglement decay. Therefore the best way to understand the qubit entanglement dynamics is by looking at the phenomena from a larger, multipartite perspective. With this work we aim at providing insight into entanglement transfer back and forth within a multipartite system subject to dissipation.

This article is structured as follows. In Sec. \ref{sec:model}, we quantitatively introduce the system of strongly driven qubits. We derive the equations of motion and present their solutions. In Sec. \ref{sec:entmeas}, we outline and discuss entanglement measures applicable to the tripartite analysis. In Sec. \ref{dissipationless} we quantify entanglement between individual subsystems in a dissipationless regime. Subsequently, in Sec. \ref{sec:entevodiss}, we study the effects of  an imperfect cavity on entanglement formation, revival and loss among different subsystems. We close the article with conclusions.

 \begin{figure}
	\centering
		\includegraphics[width=8cm]{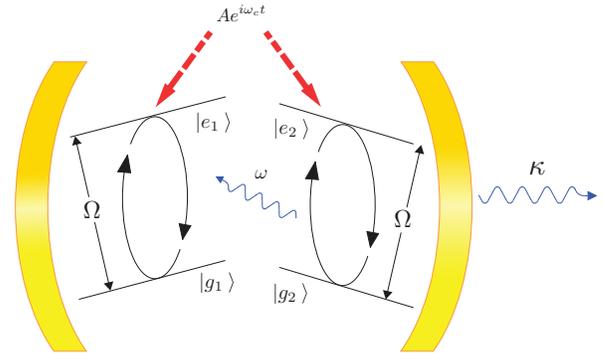}
	\caption{(Color online). Schematic representation of the setup. }
	\label{fig:cavity}
\end{figure}

\section{The Model and its Dynamics}\label{sec:model}

We consider the system of two identical qubits externally driven by a classical field of the amplitude $A$ and the frequency $\o_c$ strongly coupled to a single mode resonator \cite{2003PhRvL..90b7903S} (see Figure \ref{fig:cavity}). This system is described by the following Hamiltonian,
\bea \label{Hamiltonian}
\hat{H}&=& \hat{H}_o+\hat{H}_d+\hat{H}_I \, ,\\
\hat{H}_o&=&\frac{\O}{2} \sum\limits_{j=1}^2   \s^z_j +   \o \ahd\ah  \, ,\nnl
\hat{H}_d&=&A\sum \limits _{j=1}^2 \lr{e^{-i \o_c t} \s^+_j+e^{i \o_c t} \s^-_j} \, ,\nnl
\hat{H}_I&=& \sum\limits_{j=1}^{2} g_j\lr{\s^+_j \ah+\s^-_j \ahd} \, ,\nn
\eea
where $\O $ represents the  level spacing of each  qubit, $\o$ is the frequency of the resonator eigenmode, and $g_j$ is the coupling strength of the Jaynes-Cummings type interaction between the $j^{\rm th}$ qubit and the eigenmode of the resonator.
Additionally, we use $\s^{\pm}$ ($\ahd,\ah$) to denote the qubit (resonator) raising and lowering operators. Throughout this article we chose units where $\hbar=1$.

We work under the assumption that the qubits are driven very strongly (ensuring their greater resistance to decay $\gamma$), and are moderately coupled to the cavity mode, i.e. $A\gg \o,\o_c \gg g \gg \gamma$. Moreover, we consider the cavity dissipation rate $\k$ to be the dominant  source of decoherence in the system. A realisation of these conditions can be found for instance with superconducting qubits where $\o \approx 5$ GHz, $g \approx 100$ MHz, and $\gamma \approx 1$ MHz \cite{PhysRevA.80.043840}. 

The Hamiltonian (\ref{Hamiltonian}) is time dependent. To suppress the time dependence, one can apply a number of unitary transformations \cite{2003PhRvL..90b7903S}. First we go to the frame oscillating with the driving field frequency $\o_c$ using ${\hat{U}=\exp\lr{-i\o_c t\lr{\ahd\ah+\sum_{j}\s^z_j/2}}}$ and further to the interaction picture (IP) ${\mathcal{V}=e^{-i\lr{H_o+H_d} t} \hat{H} e^{i\lr{H_o+H_d} t}}$. Upon ignoring the quickly rotating terms $\propto e^{\pm 2 i A t}$ (the strong driving regime) the effective IP Hamiltonian  becomes (see Ref. \onlinecite{Dukalski2010} for technical details)
\bea\label{multipleHs}
\mathcal{V}= \sum\limits_{j=1}^2 g_j \s_j^x \lr{\ah e^{-i \delta t}+\ahd e^{i \delta t}}  \, ,
\eea
where we set  $\O=\o_c$ and  $\delta=\o-\o_c$. What we however see is that in the strong driving regime the coupling to the resonator no longer mediates the interaction between qubits, but is simply reduced to  the qubit state dependent bosonic displacement generator, i.e. the state of the multi-qubit system determines the state of the resonator, but the state of the resonator never affects the state of the qubits.

We model the evolution of the system with the Lindblad type master equation
\bea
\dot{\r}=\mathcal{L}\lr{\r}=-i\com{\mathcal{V}}{\r}+\k \mathcal{D}\lr{\r}\,.     \label{Lindblad}
\eea
where $\mathcal{D}\lr{\r}=\ah\r\ahd-\frac{1}{2}\acom{\ahd\ah}{\r}$ is a Markovian dissipation operator.
The system is initially in  a direct product state of a cavity field coherent state $\ket{\a}$   and a Bell state of the qubits \cite{comment1},
%
\bea
\Psi =  \frac{1}{\sqrt{2}}\lr{\ket{++}+\ket{--}}\label{psistate}\,,
\eea
where we used the diagonal  basis $\ket{\pm}=\frac{1}{\sqrt{2}}\lr{\ket{e}\pm\ket{g}}$, which are the eigenstates of the Pauli $\s_x$ matrix. The solution to the Lindblad equation (\ref{Lindblad}) is given in terms of the density operator $\r_{ij\a;kl\b}$, where the Latin indices $i$ and $k$ ($j$ and $l$) stand for the $\ket{\pm}$ state of the first (second) qubit and the Greek indices indicate the state of the cavity.
The solutions to (\ref{Lindblad}) are found in Refs. \cite{Dukalski2010,Dukalski2011}. The only non-zero entries of the density matrix $p_{ij\a;kl\b}$ with the initial condition given by (\ref{psistate}) read
{\small
\bea
p_{++\a_{+};++\a_{+}}\lr{t}    &=&   \frac{1}{2}\out{\a_+}{\a_+} \, ,\nnl
p_{++\a_{+};--\a_{-}}\lr{t}      &=&   \frac{1}{2}e^{h_1\lr{t}+i h_2\lr{t}}\out{\a_+}{\a_-} \nnl
                                              &=&   \left[ p_{--\a_{-};++\a_{+}}\lr{t} \right]^{\dagger}\, ,\nnl
p_{--\a_{-};--\a_{-}}\lr{t}        &=&   \frac{1}{2}\out{\a_-}{\a_-}\, .\label{solutions}
\eea }
Here we have defined
\bea
\a_{\pm}&=&\a e^{-kt}\pm f\lr{t}   \nnl
f\lr{t}&=&\frac{i g }{\k-  i \delta } \lr{e^{-i\delta t  }-e^{-\k t}}\nnl
h_1\lr{t}&=&
\frac{8 e^{-\k t} g^2 \k \lr{\delta  \sin\delta t  -\k \cos \delta t}}{\left(\k^2+\delta ^2\right)^2}  -\frac{8 e^{-\k t} g^2 \k^2 \cos \delta t }{\left(\k^2+\delta ^2\right)^2}\nnl
&&-\frac{4 g^2 \k t}{\k^2+\delta ^2} +\frac{2 e^{-2 \k t} g^2}{\k^2+\delta ^2} +\frac{2 g^2 \left(3 \k^2-\delta ^2\right)}{\left(\k^2+\delta ^2\right)^2}\,, \label{h1}\\
h_2\lr{t}&=&h_2\lr{0}-\frac{2 e^{-2 \k t} g (\k \a_r+\a_i \delta )}{\k^2+\delta ^2} \nnl &&\hspace{-1cm}-\frac{2 e^{-\k t} g \lr{\lr{\a_i \delta -3 \k \a_r} \cos \delta t+\lr{3 \k \a_i+\a_r \delta } \sin \delta t }}{\k^2+\delta ^2}\,,\nn
\eea
where $\a_{r,i}$ are the real and imaginary parts of the initial state of the cavity and $g=g_1+g_2$ is the effective qubits-cavity coupling.

The coherent state considered here has a continuous, time dependent amplitude. Such state is represented by a vector spanning the whole of the infinite Fock space, making the qubits-cavity system $2\times 2\times \infty$ dimensional rendering some of the entanglement measures inapplicable. These different coherent states however can be written in bases found in \cite{ZhangXu}. Using the fact that every coherent state is a single-parameter state, we can recast the two coherent states $\ket{\a_{\pm}}$ in a two-dimensional form  by means of orthogonalisation through the Gram-Schmidt process.
\bea
\ket{\uparrow}&=&\ket{\a_{+}}\,,\nonumber\\
\ket{\downarrow}&=&\frac{1}{\sqrt{1-\lra{\chi}^2}}\lr{\ket{\a_{-} }-\chi \ket{\a _{+} }}\, .\nonumber\\
\chi\lr{t}&=&\inner{\a_{+}}{\a_{-}}\label{lambda}\,.
\eea
such that $\inner{\uparrow}{\downarrow}=0$ and where the inverse transformation reads
\bea
\ket{\a_{+}}&=&\ket{\uparrow}\,,\nonumber\\
\ket{\a_{-}}&=&\sqrt{1-\lra{\chi}^2}\ket{\downarrow}+\chi \ket{\uparrow}\, .\nnl 
\eea
As a result the system is now reduced to $2\times 2\times 2=8$ dimensions. In the last subspace the bases definitions are time dependent, but the resulting set of states $\left\{\ket{\uparrow},\ket{\downarrow}\right\}$ is orthogonal at any point in time. In this form, we can easily use the established entanglement formalism.

The full system time-dependent density operator $\r$ in the $\mathbb{C}^2\times \mathbb{C}^2\times \mathbb{C}^2$ space spanned by the qubits diagonal basis and the orthogonalised coherent state basis reads
\bea\label{rho8by8}
\rho=\lr{\begin{array}{ccc}
 K             & \cdots & L \\
\vdots         & \ddots & \vdots\\
 L^{\dagger}   & \cdots & M
\end{array}}\,,
\eea
which is a sparse $8\times 8$ matrix where the only non zero elements are contained in the $2 \times 2$ blocks
\bea
K &=&\frac{1}{2}\lr{\s_z+I_2}=\lr{\begin{array}{ll}
 1 & 0 \\
 0 & 0
\end{array}}\,,\nonumber\\
L &=&\lr{\begin{array}{ll}
 e^{h_1+i h_2 } \chi  & e^{h_1+i h_2 } \sqrt{1-\lra{\chi }^2} \\
 0 & 0
\end{array}}\,,\nonumber\\
M &=&\frac{1}{2}\lr{\begin{array}{ll}
 \lra{\chi }^2   & \chi  \sqrt{1-\lra{\chi }^2} \\
 \chi^*  \sqrt{1-\lra{\chi }^2} & 1-\lra{\chi }^2
\end{array}}\,.\nonumber
\eea

\section{Entanglement Measures}\label{sec:entmeas}
It is not always easy to establish non-separability (entanglement) of a system based on the form of a density operator. The factors that play a role here are, among others, bases choice and dimensionality of the system. Fortunately, the last two decades brought developments in the field of entanglement measures \cite{wootters,Coffman2000, vidal, Horodecki1996}. There it was shown that stepping beyond $2\times 2$ physical systems, where one oftentimes uses concurrence, one has to account for a greater number of correlations between individual players in a multipartite physical system \cite{Guhne20091} and can choose from entanglement witnesses, negativity, or the three-tangle. We devote this section to briefly review some of the most important aspects which we will find useful in out subsequent analysis.

One of the first entanglement measures to be introduced and since then widely used for $2\times 2$ dimensional systems is concurrence \cite{wootters}. Its mathematical form given by
\bea
\mathcal{C}={\rm max} \lr{0,\sqrt{\lambda_1}-\sqrt{\lambda_2}-\sqrt{\lambda_3}-\sqrt{\lambda_4}}\, ,\label{concurrence}
\eea
where $\lambda_i$ are the  eigenvalues 
of  ${R=\lr{\s_y\otimes\s_y}\r^*\lr{\s_y\otimes\s_y}\r}$, with $\lambda_1$ being the largest of them, $\s_y$ being the Pauli $y$-matrix. The value of $\mathcal{C}$ ranges from zero (no entanglement) to one (maximum entanglement). This measure however no longer suffices when dealing with systems involving more than two two-dimensional subsystems.

In order to study entanglement in the tripartite system, we use Horodeckis' separability criterion \cite{Horodecki1996} and stemming from it  negativity \cite{vidal}  to quantify tripartite entanglement. Using the partial transposition (in the second subspace)  defined by
\bea
\r&=&\sum\limits_{ijkl}\a_{ijkl} \out{i}{j}\otimes\out{k}{l}\, ,\nnl
\r^{pT_2}&=&\sum\limits_{ijkl}\a_{ijkl} \out{i}{j}\otimes\out{l}{k}\,,\nn
\eea
this criterion states that the density operator of an entangled state upon transposition in one of the subspaces will have at least one negative eigenvalue.  Negativity is then the sum of absolute values of negative eigenvalues of $\r^{pT}$.

Thus when studying a tripartite system composed of three subsystems $A$, $B$ and $C$ (in this case $A$ and $B$ are the qubits and $C$ is the cavity, but the labeling is completely arbitrary), we can find the degree of entanglement between the combined bipartite subsystem $AB$ and subsystem $C$, by partial transposing the density operator $\r_{ABC}$ of the system in the basis states that span the subsystem $C$, and later adding up all of the absolute values of the negative eigenvalues of $\r_{ABC}^{pT_C}$. As a result we obtain negativity ${\rm Neg}\lr{AB|C}$, which when equal to zero corresponds to no (or bound i.e. a state with zero negativity that is not separable) entanglement and when equal to $\frac{1}{2}$ indicates maximum bipartite entanglement. To get the full picture of tripartite entanglement in this system we need to also calculate ${\rm Neg}\lr{AC|B}$ and ${\rm Neg}\lr{BC|A}$, where the partial transposition is made in the subsystem $B$ and $A$ basis respectively.

 Since the dimension of this system is larger than six, i.e. the limit imposed by the Horodeskis' separability criterion, we could encounter bound entanglement . We could avoid this subtlety by creating a map  $\mathbb{C}^2\otimes \mathbb{C}^2\to \mathbb{C}^2 $ which maps the entangled state $\ket{++}+\ket{--}$ onto a superposition $\ket{+}+\ket{-}$ reducing the dimensionality of the system, by removal of permanently empty rows and columns of the density operator. This will however prove to be unnecessary, as we will see later, that the only time when negativity is strictly zero is at the expected periodically distributed points in time $\delta t=2\pi n$ for integer $n$, when the qubits are completely disentangled from the cavity (see Figure \ref{fig:gr1} and Eq. \ref{solutions}); something that can be easily seen without invoking any entanglement measures formalism. Thus the excessive dimensionality of our system posses no problems with regards to using negativity as an entanglement measure.

One drawback of the negativity is that it only provides information about entanglement of two parts of the system under partitioning of our choice and does not tell us anything about the total entanglement present. Adapting the approach of \cite{Yu2004} we can use the sum of the bipartite entanglements
\bea
TE={\rm Neg} \lr{Q_1 C| Q_2}+{\rm Neg} \lr{Q_2 C| Q_1}+{\rm Neg} \lr{Q_1 Q_2| C} \, .\nnl\label{TEeq}
\eea
where we replace the arithmetic mean by a direct sum so that we can use $TE=1$ as an easier reference point for how much entanglement was there initially in the system. As a result of this definition we get that  $0\leq TE \leq \frac{3}{2}$, where the lower bound indicates no  and the upper bound indicates maximal tripartite entanglement, that of for example the $GHZ_3$ state
$$
\ket{GHZ_3}=\frac{1}{\sqrt{2}}\lr{\ket{000}+\ket{111}}\, ,
$$
where all negativities ${\rm Neg} \lr{Q_1Q_2|Q_3}={\rm Neg}\lr{Q_1Q_3|Q_2}={\rm Neg}\lr{Q_2Q_3|Q_1}=\frac{1}{2}$.

The GHZ$_3$ state shows a feature that will be important to our further discussion, namely tripartite entanglement sharing. In this state (as opposed to the $W$-state) the individual subsystems share bipartite entanglement but partial tracing over one of the subsystem (loosing a qubit) results in a statistically mixed state (the $W$ state results in a Bell state).

This result is known as the monogamy of entanglement which states that a subsystem $A$  maximally entangled to second subsystem $B$ cannot simultaneously be entangled with another subsystem $C$. This has been first formulated by the Coffman, Kundu and Wootters \cite{Coffman2000} in terms of the inequality
\bea
\mathcal{C}_{\lr{A|BC}}^2\geq \mathcal{C}_{\lr{AB}}^2+\mathcal{C}_{\lr{AC}}^2\, ,\label{CKWineq}
\eea
where $\mathcal{C}^2$ are the tangles (concurrences squared). Here $\mathcal{C}_{\lr{A|BC}}$ is found by reducing the dimensionality of the density operator $\r$ down to the subspace spanned by the two eigenvectors of $\r'$  with non-zero eigenvalues of the $BC$ subspace and $\mathcal{C}_{\lr{AB}}$ and $\mathcal{C}_{\lr{AC}}$ are concurrences of bipartite subsystem obtained by partial tracing the total tripartite system over the subsystem $C$ and $B$ respectively.

The inequality (\ref{CKWineq}) can be used to define a three-tangle given by the inequality mismatch
$$
\tau_{ABC}=\mathcal{C}_{\lr{A|BC}}^2- \mathcal{C}_{\lr{AB}}^2-\mathcal{C}_{\lr{AC}}^2\,.
$$
This new quantity tells us how much of residual tripartite entanglement is there when all of the bipartite contributions are taken away. It is easy to see from the definition of concurrence that the three-tangle ranges from zero (no shared entanglement) to 1 (completely inseparable state of ${\rm GHZ}_3$ type).

From the solutions to the equation (\ref{Lindblad}) we see that the qubits entangle with the cavity, which for $\delta,\k\to 0$ and $t\to\infty$ leads to a perfectly entangled GHZ-like state, as the coherent state amplitudes undergo a shift in opposite directions,
\bea
\ket{\Psi,\a}\lr{0}&=&\ket{\Psi}\otimes\ket{\a_0}\, ,\nnl
\ket{\Psi,\a}\lr{t}&=&\frac{1}{\sqrt{2}}\lr{\ket{++}\ket{\a_0-2 i g t}+\ket{--}\ket{\a_0+2 i g t}}\nnl
&&\equiv\frac{1}{\sqrt{2}}\lr{\ket{++,A_+} +\ket{--,A_-}}\nn \ .
\eea
This state is different from the GHZ-state since $\inner{A_+}{A_-}\neq 0$. Upon taking the partial trace of $\r\lr{t}$ over the cavity, we observe that the diagonal entries of the density operator $\pTr{c}{\out{\Psi}{\Psi}}$ are unchanged, but the entries $\out{++}{--}$ acquire time dependence $e^{-2g^2 t^2}$ which mimics the dephasing of the two qubit state. This is because as $t$ grows the state of the system more and more closely resembles the GHZ-state, and taking the trace leaves  the state in a completely mixed state to a larger extent. It is a continuous in time analogue of the GHZ state formation from the original Bell state.

The effect of entanglement revival in this system is brought about by the presence detuning between the cavity and the resonantly driven qubits. Since we would be interested in periodic revival of entanglement for the remainder of out analysis we have to keep $\delta\neq 0$.
In what follows we will mainly focus on the dissipationless case as it provides a very good insight into qualitative as well as quantitative aspects of qubits-cavity entanglement dynamics. Later we will study the effect  a combination of dissipation and detuning on the inter-qubit as well as  the qubits-cavity entanglement.

\section{Dissipationless cavities}\label{dissipationless}
In the closed system (in the $\k \to 0$ limit ), the non-resonant interaction between the qubits and the cavity will result in formation of a coherent state with an amplitude oscillating in time with frequency $\delta$. Under these conditions the complete state of the system is still represented by Eq. \ref{solutions} where the previously defined expressions are replaced by
\bea
h_1&=&0\,,\label{xdef2}\\
\chi&=&e^{  g \lr{\cos  \delta t-1}  \lr{4  \frac{ g }{\delta} -2 i b   +2 i a \lr{\sin \delta t}{\cos  \delta t-1} }/\delta }\nonumber\,  \ ,
\eea
where $a$ and $b$ are the real and imaginary parts of the initial coherent state amplitude, and the value of $h_2\lr{t}$ will have no effect on the result.  Upon partial transposing expression (\ref{rho8by8}) with respect to the cavity subspace we get
$$
\rho^{pT_C}=\lr{\begin{array}{ccc}
 K             & \cdots & L^T \\
\vdots         & \ddots & \vdots\\
 L^{*}         & \cdots & M^T
\end{array}}
$$
where there is  only one negative eigenvalue, and the negativity takes the form
{\small
\bea
{\rm Neg} \lr{Q_1 Q_2| C}=\frac{1}{2} \lr{1-e^{\frac{8 g ^2 \lr{\cos   \delta t -1}}{\delta ^2}}}^{\frac{1}{2}}\label{NegQQC}\,.
\eea}
Taking the partial transposes in the individual qubit spaces, we get a symmetric result
{\small
 \bea
{\rm Neg} \lr{Q_1 C| Q_2}\lr{t}={\rm Neg} \lr{Q_2 C| Q_1}\lr{t}=\frac{1}{2}\,.
\eea}
Note that the initial state of the cavity $\a$ has no effect on the results. 
It is important to note that under dissipationless evolution the entanglement between the two subsystems spanned by the joint qubit-cavity subspace and the other qubit does not change with time (i.e. there will be no bipartite entanglement variation between the two qubits), thus since ${\rm Neg} \lr{Q_1 Q_2| C}\geq 0$ the total entanglement can only increase relative to its initial value.


\begin{figure}
	\centering
		\includegraphics[width=8cm]{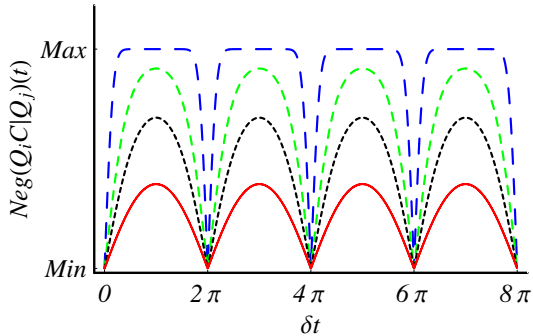}
		\caption{(Color Online) Negativities: qubits vs cavity ${\rm Neg} \lr{Q_1 Q_2| C}$   for a  qubits initiated in the $\ket{\Psi}$ state (\ref{psistate}).   Plots made for $g_1=g_2=100$MHz and $\delta=\{0.1,0.3,0.5,1\}$ GHz in blue, green, black, red (increasing dashing frequency) respectively. The peak of the maximum is always where $\delta t$ takes values of odd integers of $\pi$ and their value is $\frac{1}{2} \lr{1-e^{-\frac{16 \lr{g_1+g_2}^2}{\delta ^2}}}^{\frac{1}{2}}$}
	\label{fig:gr1}
\end{figure}



The behaviour of ${\rm Neg} \lr{Q_1 Q_2| C}$ (see Fig. \ref{fig:gr1}) displays periods of entanglement and disentanglement between the qubits and the cavity. This is due to the fact that every period of length $2\pi/\delta$, the coherent state of the cavity returns to its initial state. Figure \ref{fig:gr1} also shows that the strength of qubits-cavity entanglement formed depends on detuning. The coherent states under detuned driving of the qubits  change their amplitudes to a limited extent. The values of the coherent state amplitude and phase follow a circular trajectory in a complex plane centered at $\a_o\pm g/\delta$ with periods $\delta$ and radii $g/\delta$, where $\a_o$ is the initial coherent state amplitude.

\begin{figure*}
	\centering
		\includegraphics[width=8cm]{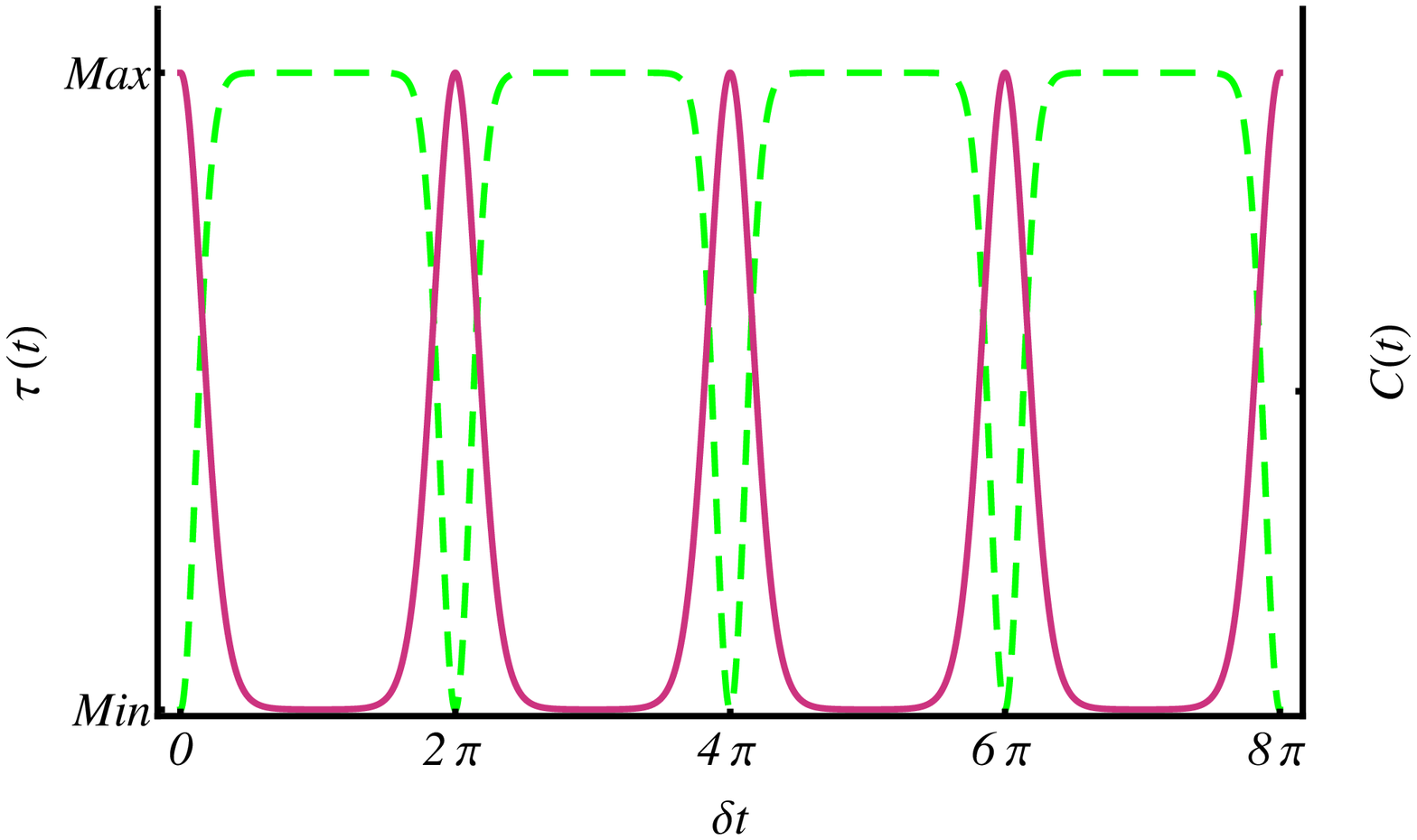}
		\includegraphics[width=8cm]{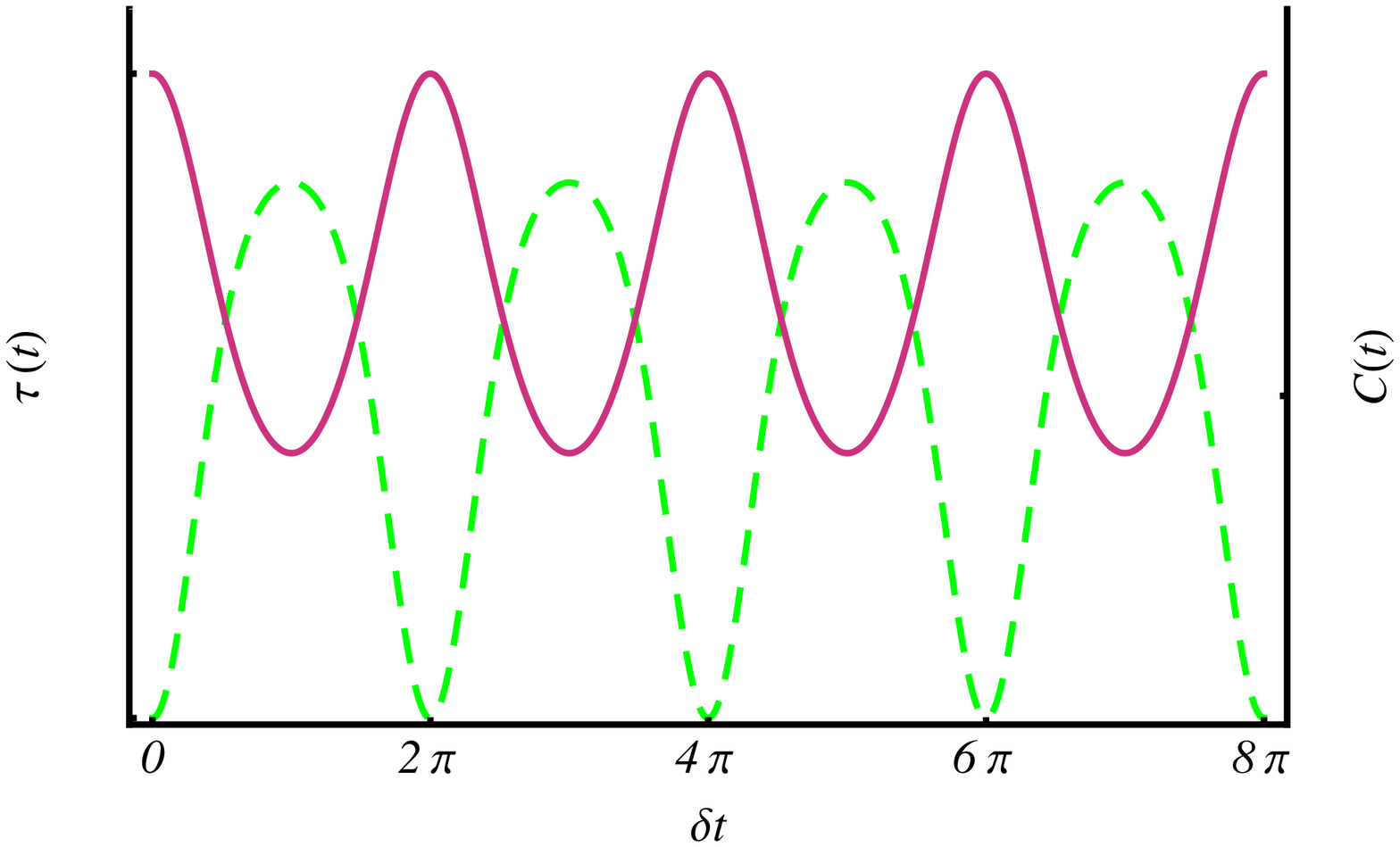}
	\caption{(Color Online) A plot of a three-tangle and qubit-qubit subsystem concurrence for a dissipationless cavity (plum solid and green dashed lines respectively). We see that the at times of decreased qubit-qubit entanglement, the three-tangle grows, supporting our conjecture that the qubit-qubit entanglement is transferred to entanglement of a tripartite system. Plots made for $g_1=g_2=100$ MHz, and $\delta=200$ MHz (left) and $\delta=600 MHz$ (right)}
	\label{fig:tanglePlot}
\end{figure*}

The creation of entanglement between the qubits and the cavity, however bares consequences to the qubit-qubit subsystem. Previously in Ref. \cite{Dukalski2010,Dukalski2011} we saw that qubits can undergo oscillations in their relative entanglement strengths (even if we take the  $\k\to 0$ limit of equation (10)).
By considering the solutions and  Figure \ref{fig:gr1} we can see that throughout the evolution the qubit-qubit subsystem oscillates between completely entangled and partially mixed states,
\bea
\Tr{\r}_c=\out{\Phi}{\Phi}\leftrightarrow \r_{Q_1Q_2}\,,\nnl
  \r_{Q_1Q_2}=\frac{1}{2} \lr{\begin{array}{llll}
 1      & 0  & 0  & \epsilon\lr{t} \\
 0      & 0  & 0  & 0  \\
 0      & 0  & 0  & 0  \\
 \epsilon\lr{t}^{*}  & 0 & 0  & 1
\end{array}} \,, \nonumber
\eea
where $\Tr{\r}_c$ denotes a partial trace over the cavity states and $\epsilon\lr{t}= \inner{\a_+}{\a_-}$.
This has to do only with the fact that the $\Tr{p_{++;--}}_c$ entries carry time dependence, while the populations i.e. the $\Tr{p_{++;++}}_c$ and $\Tr{p_{--;--}}_c$ entries, are constant in time.  This should not be surprising as the effective Hamiltonian does not allow for the individual state populations to change.
 \begin{figure*}[t]
	\centering
		\includegraphics[width=5.5cm]{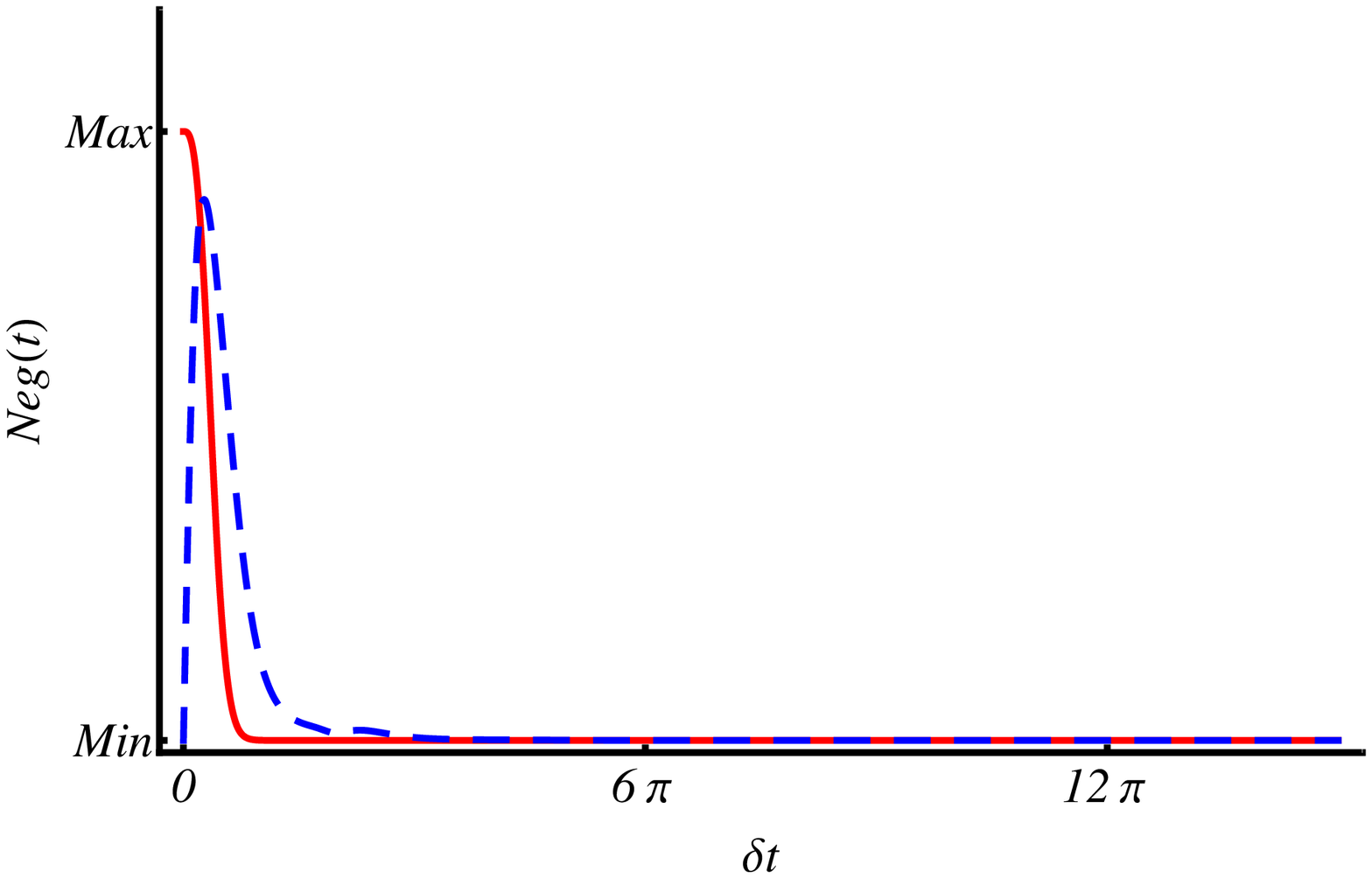}
    \includegraphics[width=5.5cm]{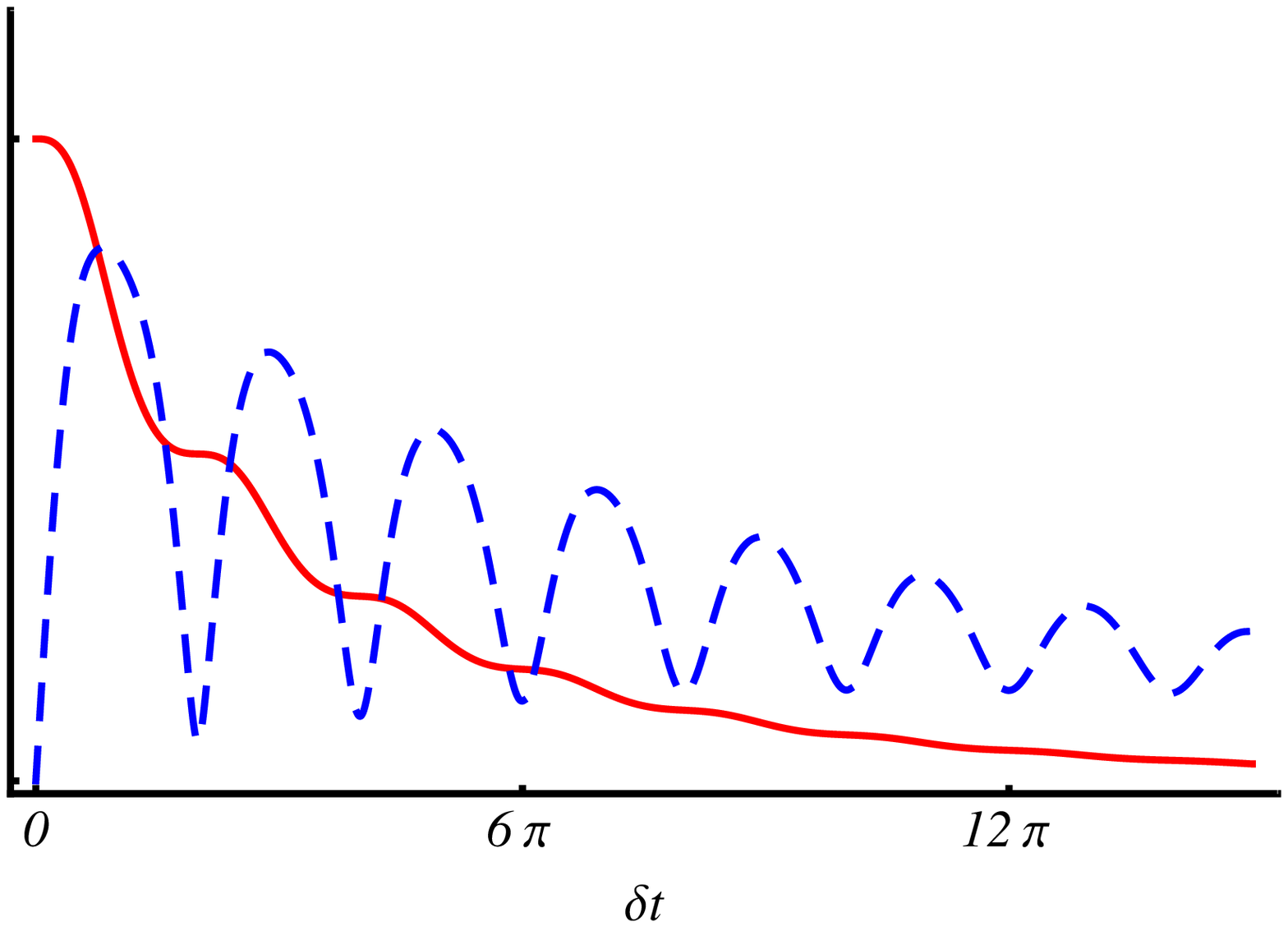}
    \includegraphics[width=5.5cm]{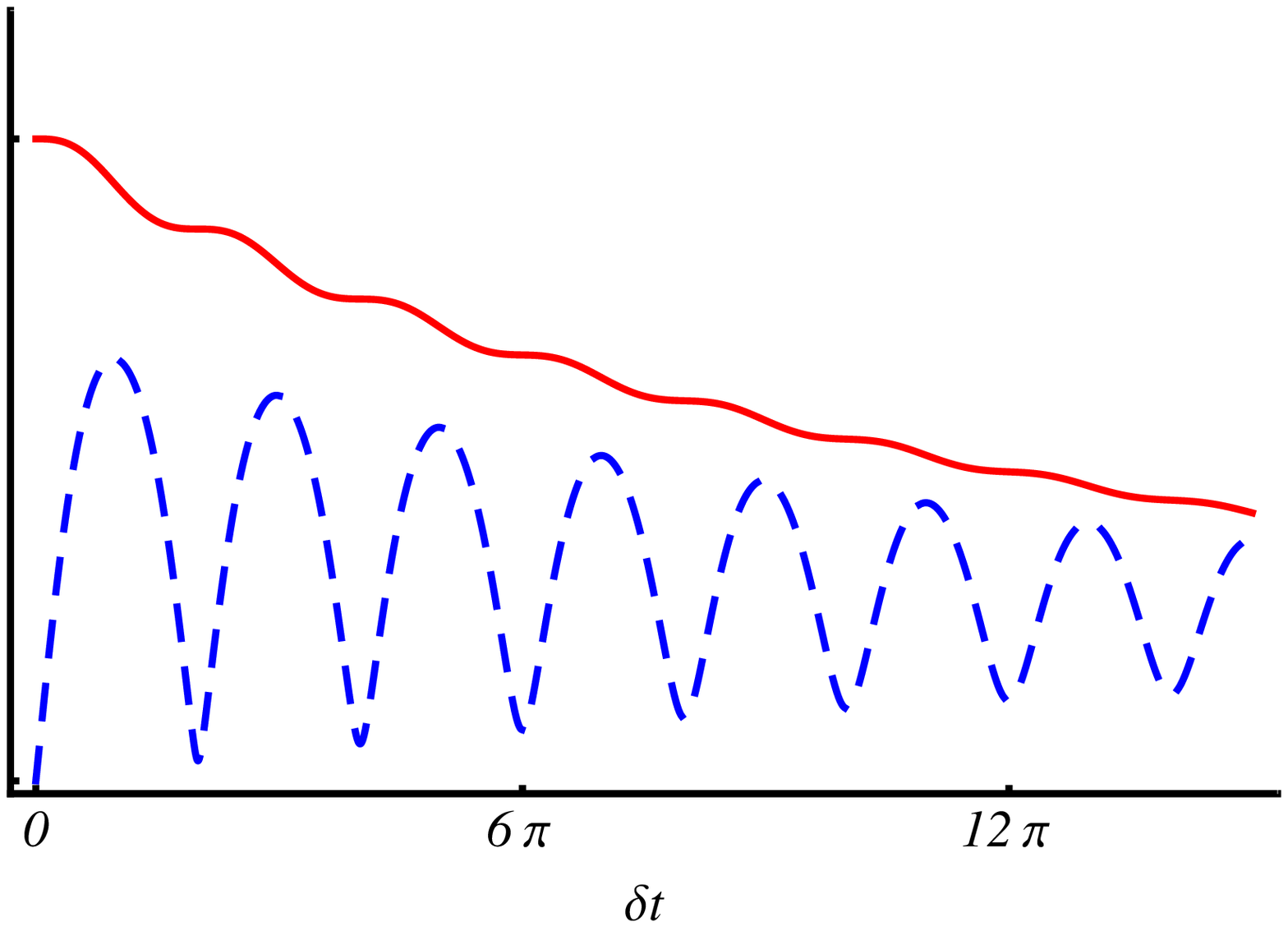}\\
		\includegraphics[width=5.5cm]{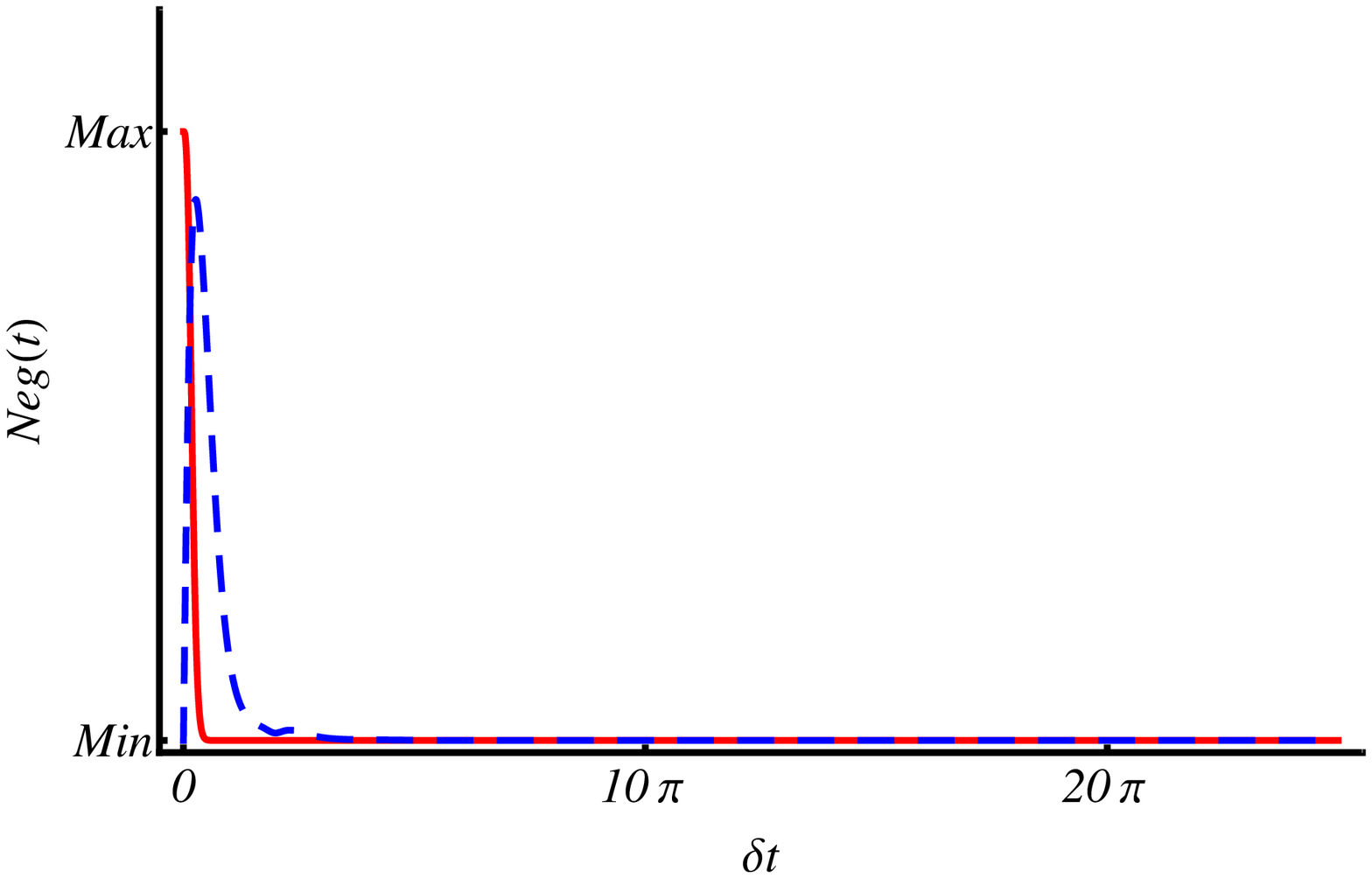}
		\includegraphics[width=5.5cm]{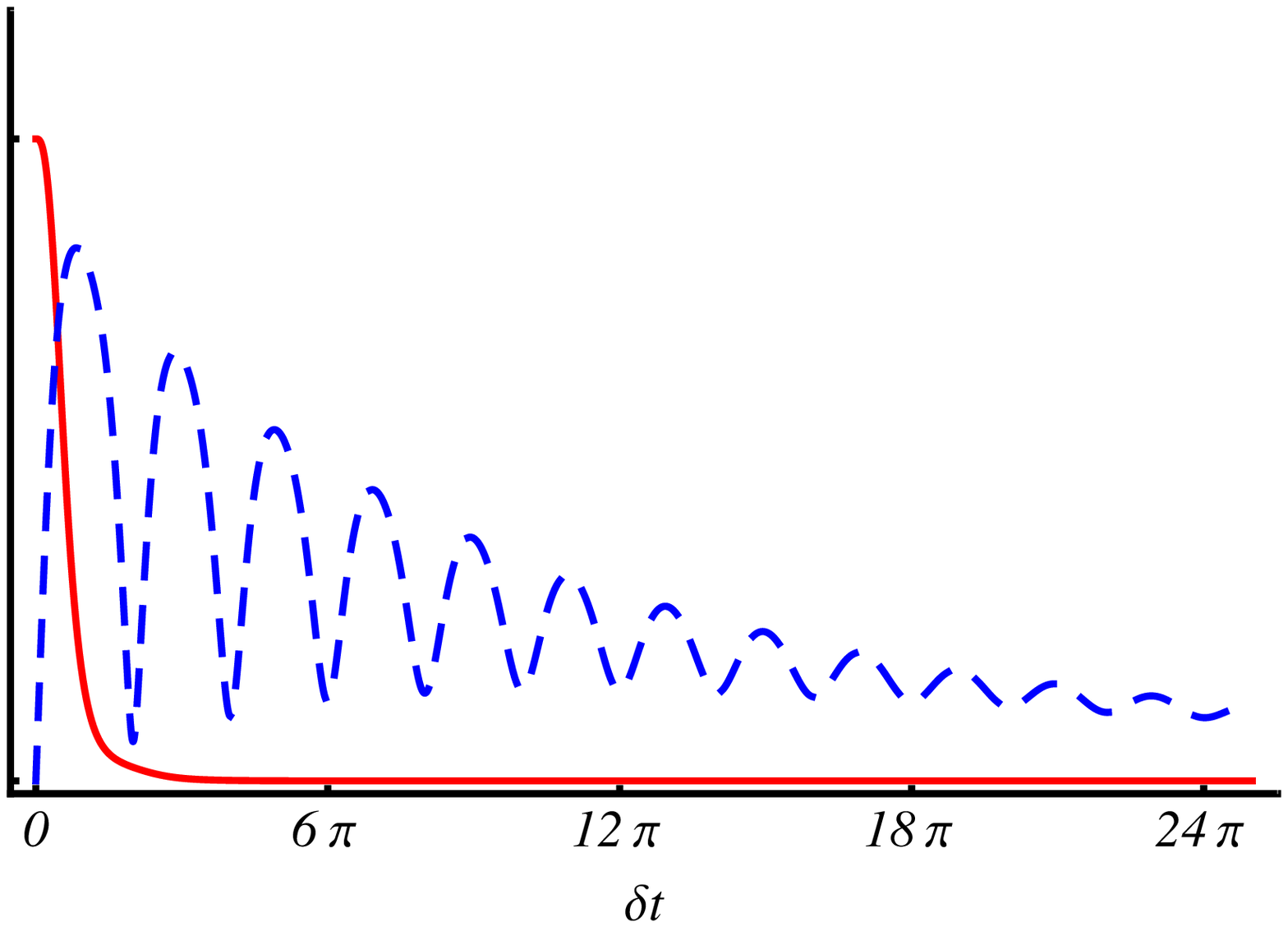}
		\includegraphics[width=5.5cm]{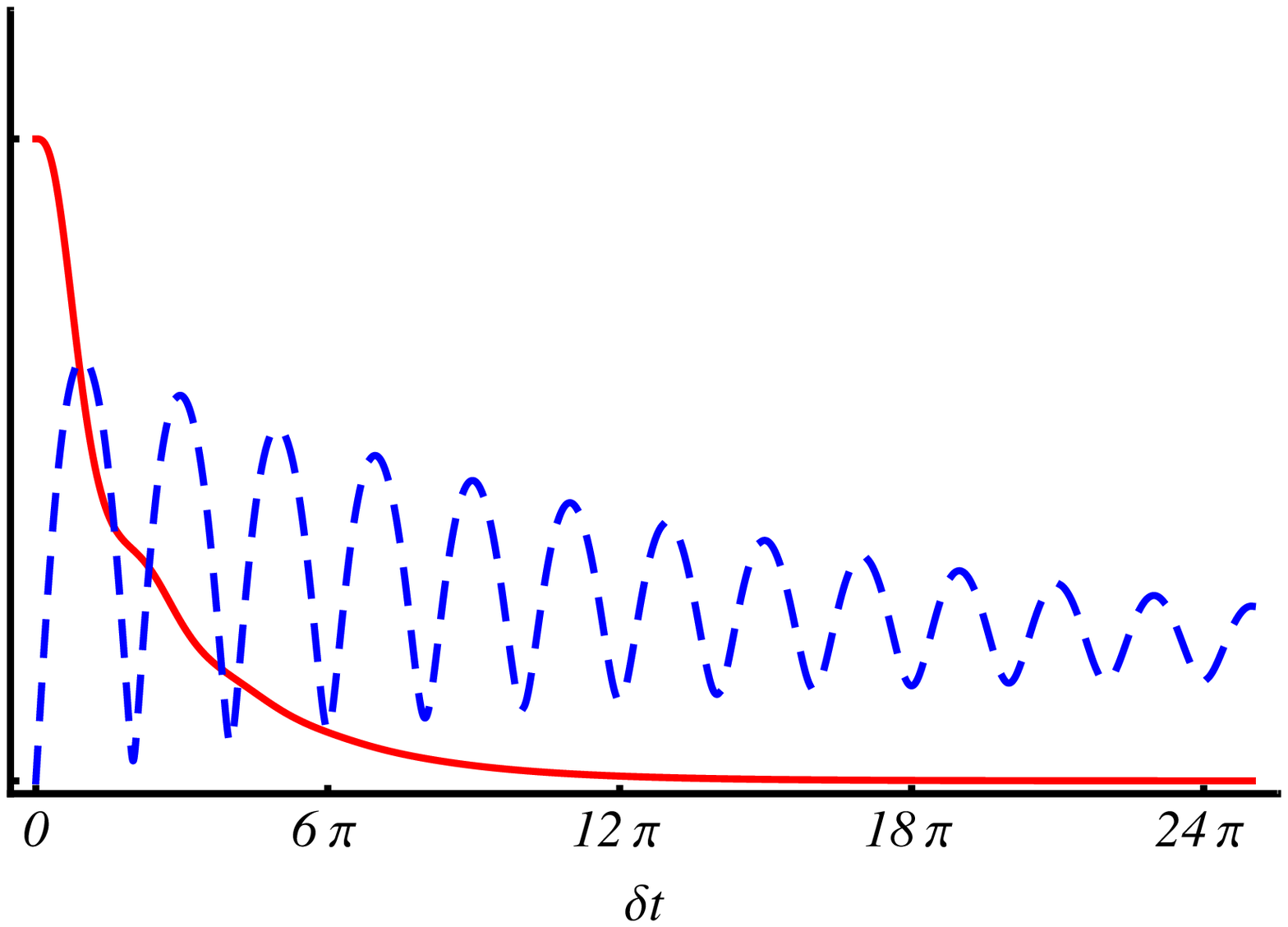}\\
		\includegraphics[width=7.5cm]{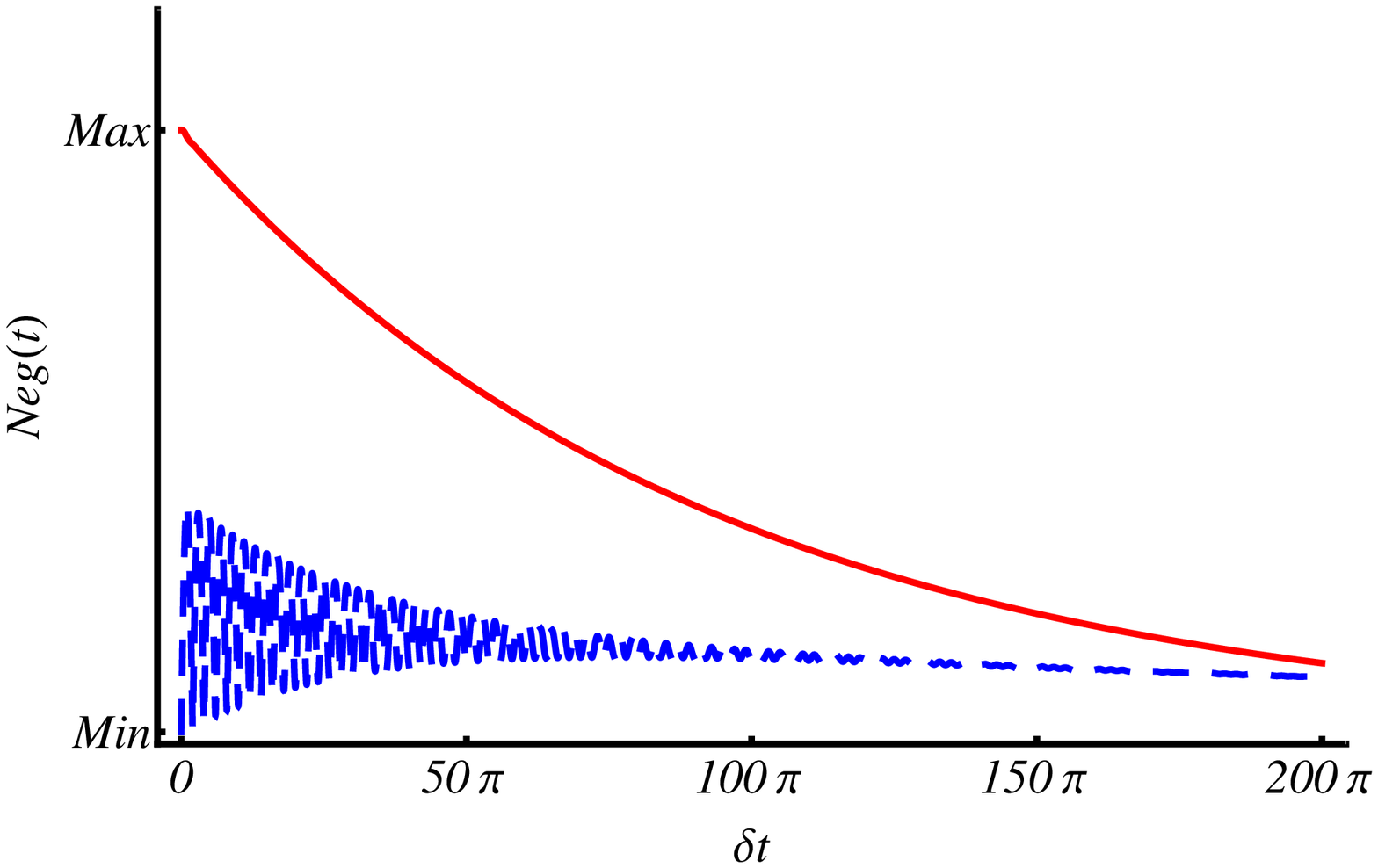}
		\includegraphics[width=7.5cm]{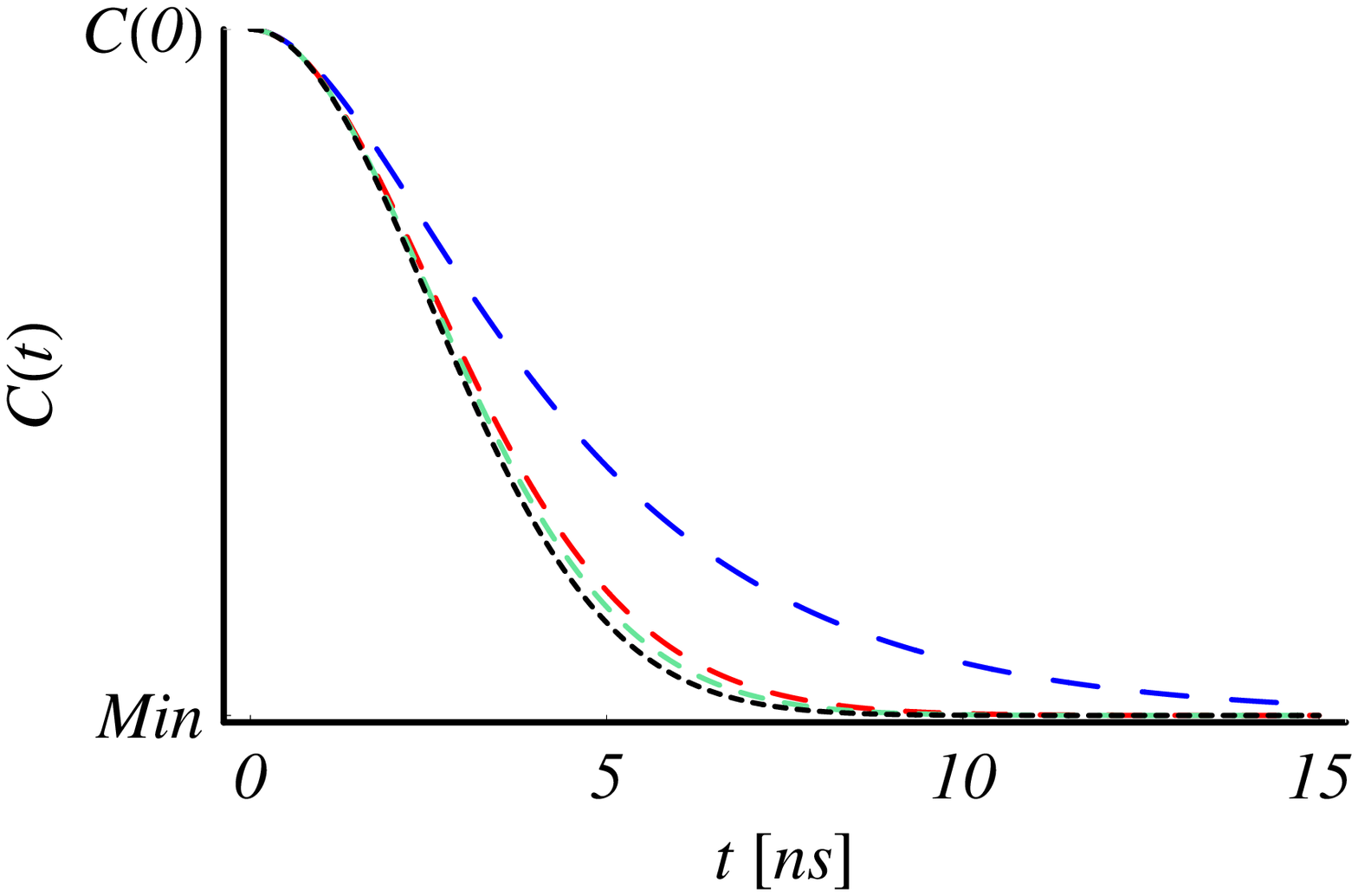}
	\caption{(Color online)   The plot of ${\rm Neg} \lr{Q_1 C| Q_2}\lr{t}$ (red-solid) and ${\rm Neg} \lr{Q_1 Q_2 |C}\lr{t}$ (blue-dashed) made for $g_1+g_2=200$ MHz, $\delta=100$ MHz (first column), $\delta=300$ MHz (second column) and $\delta=500$ MHz (third column), and $\k=100$ MHz (first row) and $\k=1$ MHz (second row).  Bottom-left: ($g=50$ MHz, $\delta=1$ GHz, $\k=400$ MHz) we see that the ${\rm Neg} \lr{Q_1 Q_2 |C}\lr{t}$ is much longer lived and eventually decays but always later than ${\rm Neg} \lr{Q_1 C| Q_2}\lr{t}$.\\
We see that the entanglement generated between the cavity and the two qubits depends on a delicate balance between the detuning and the dissipation rate.
Bottom-right: comparison between the qubit-qubit concurrence decay rate with and without decay. Plots made for $g_1+g_2=200$ MHz, $\k=\{500,100,50,0\}$ MHz in blue, red, green, black (increasing dashing frequency) respectively. We see that larger cavity decay rates offer slower concurrence decays, however due to further cavity-environment coupling asymptotically concurrence is zero.}
	\label{fig:new2}
\end{figure*}

When just the two qubits are considered (trace over the cavity),  the  entanglement between them will undergo fluctuations. In line with the monogamy of entanglement every time when the cavity almost completely entangles to the qubits, the qubits themselves must share very little bipartite entanglement, because now the system forms a tripartite entangled state as a whole.

When we calculate the amount of tripartite-shared entanglement we find
$$
\mathcal{C}_{\lr{C|Q_1Q_2}}=\sqrt{1-\exp\lr{\frac{8 g^2 (\cos  \delta t-1)}{\delta ^2}}}\,,
$$
which incidentally is just twice the negativity in Eq. (\ref{NegQQC}) \cite{eisert1999}. If we  calculate the $\mathcal{C}_{\lr{C Q_1}}$ ($\mathcal{C}_{\lr{CQ_2}}$) by taking the trace over the subsystem $Q_2$ ($Q_1$), we find them equal to zero. Thus we get that
$$
\tau_{CQ_1Q_2}=1-\exp\lr{\frac{8 g^2 (\cos  \delta t-1)}{\delta ^2}}
$$
Thus we see that for small detunings most of the entanglement is shared among the three entities and only when time $t$ is close to an integer multiples of $2\pi/\delta$ then the entanglement between the qubits subsystem and the cavity is lost, resulting in recovery of the bipartite entanglement between the qubits (see Figure \ref{fig:tanglePlot}).

The total entanglement $TE$ will simply be ${\rm Neg} \lr{Q_1 Q_2| C}+1$, thus we can extend the conclusions above to the total entanglement in the system as their qualitative nature does not change. It is interesting to analyse the case when dissipation is present, which is the focus of the next section.


\begin{figure*}[t]
	\centering
		\includegraphics[width=0.40\textwidth]{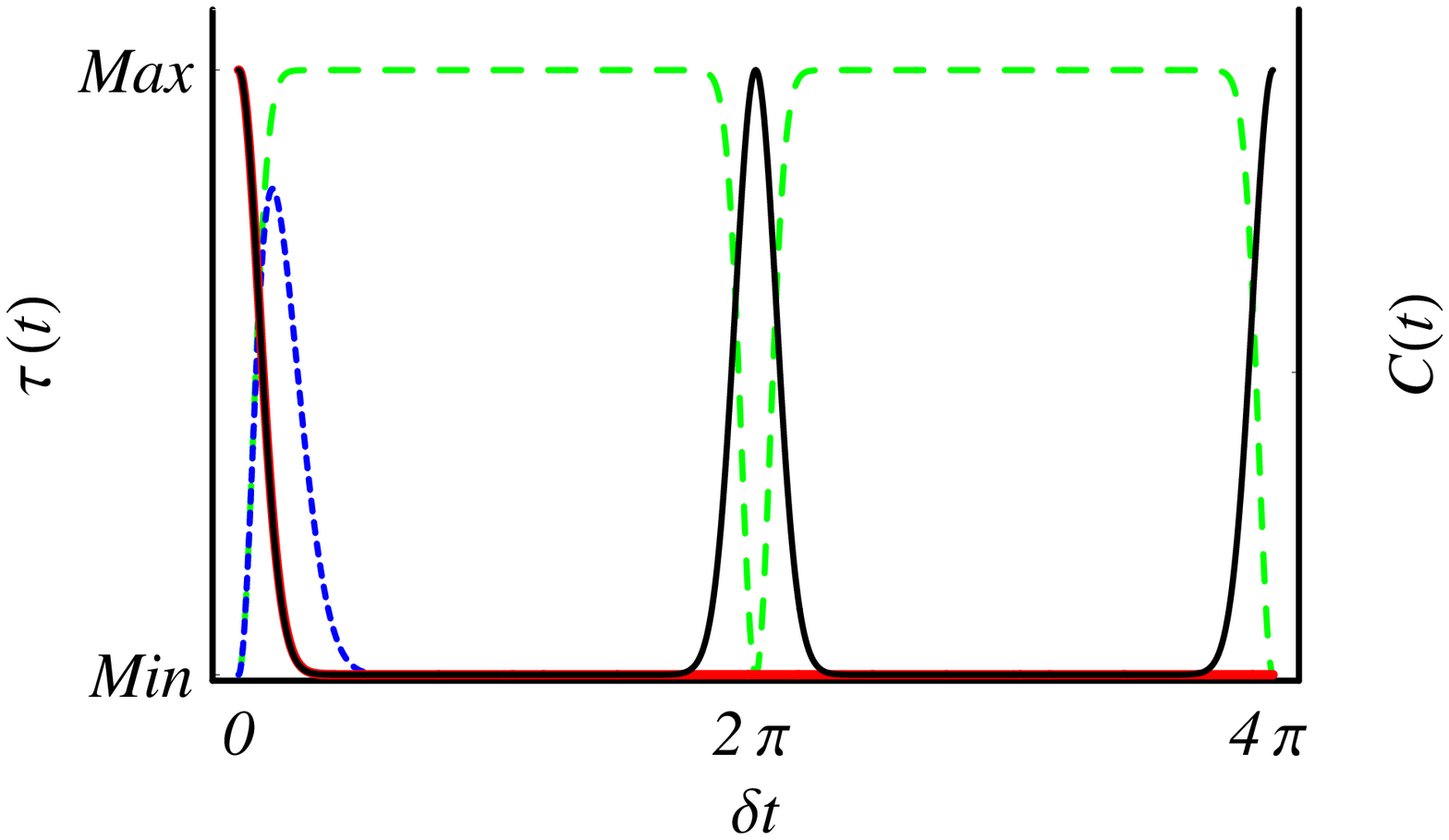}
    \includegraphics[width=0.40\textwidth]{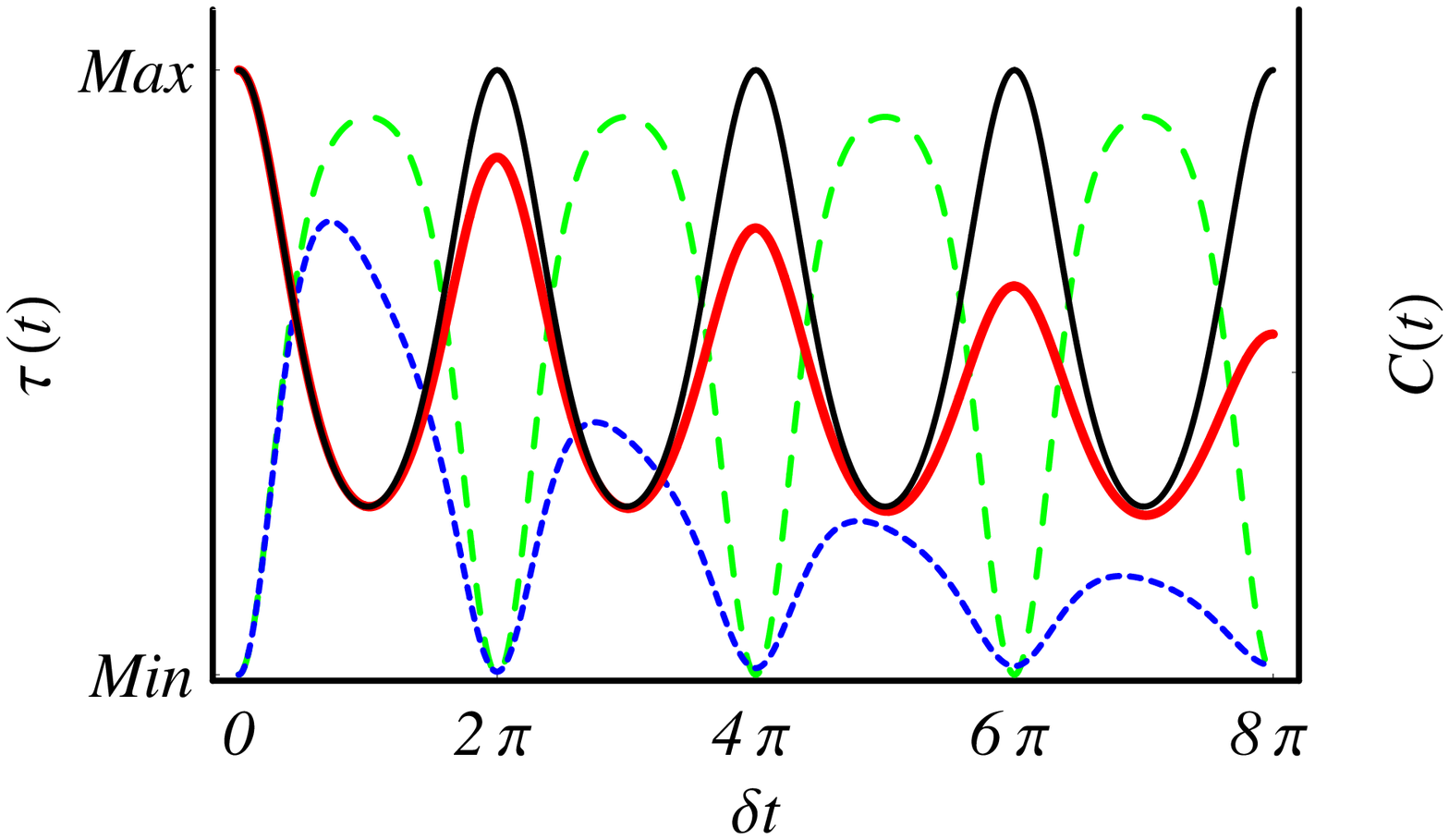}\\
    \includegraphics[width=0.40\textwidth]{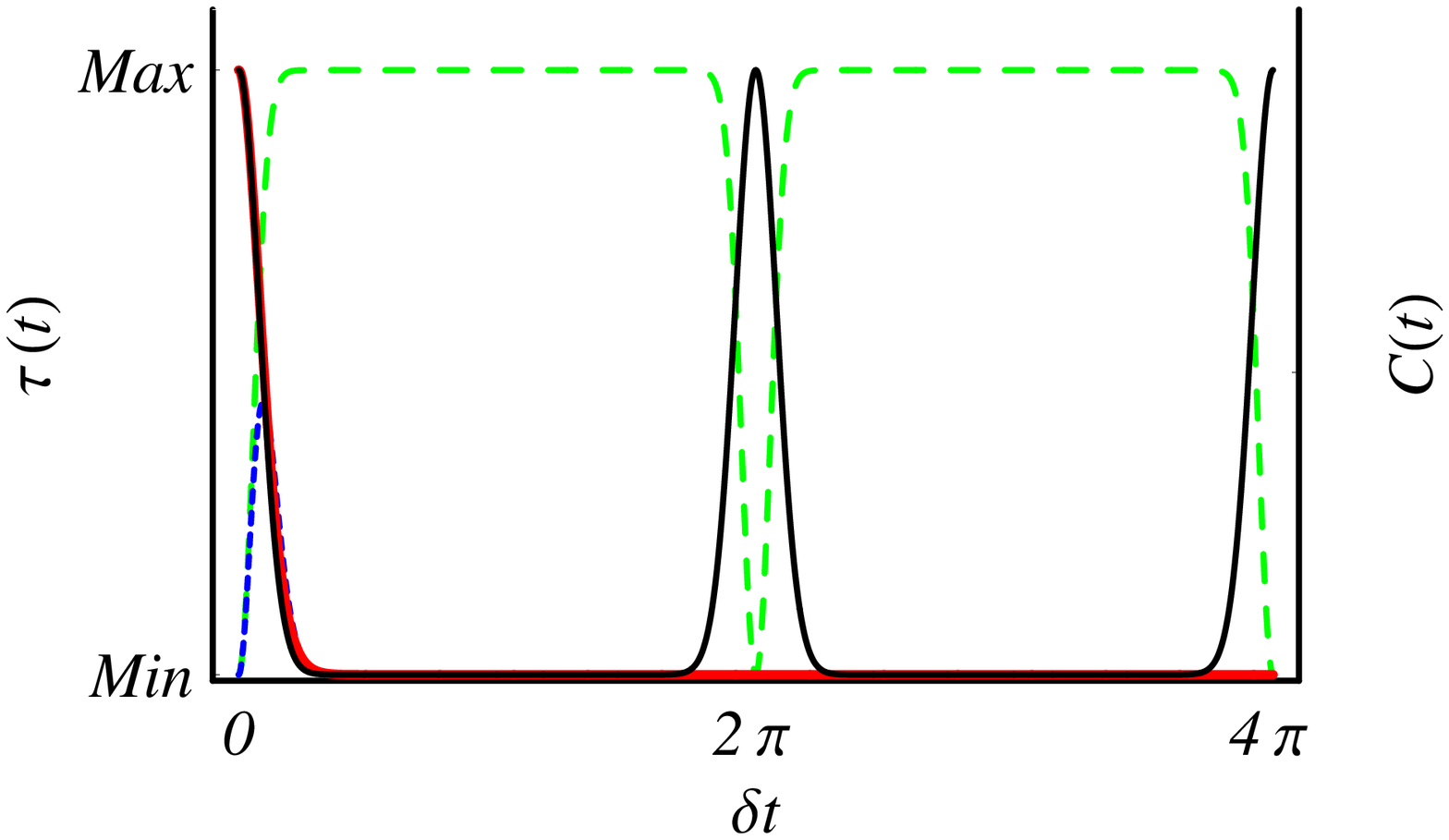}
		\includegraphics[width=0.40\textwidth]{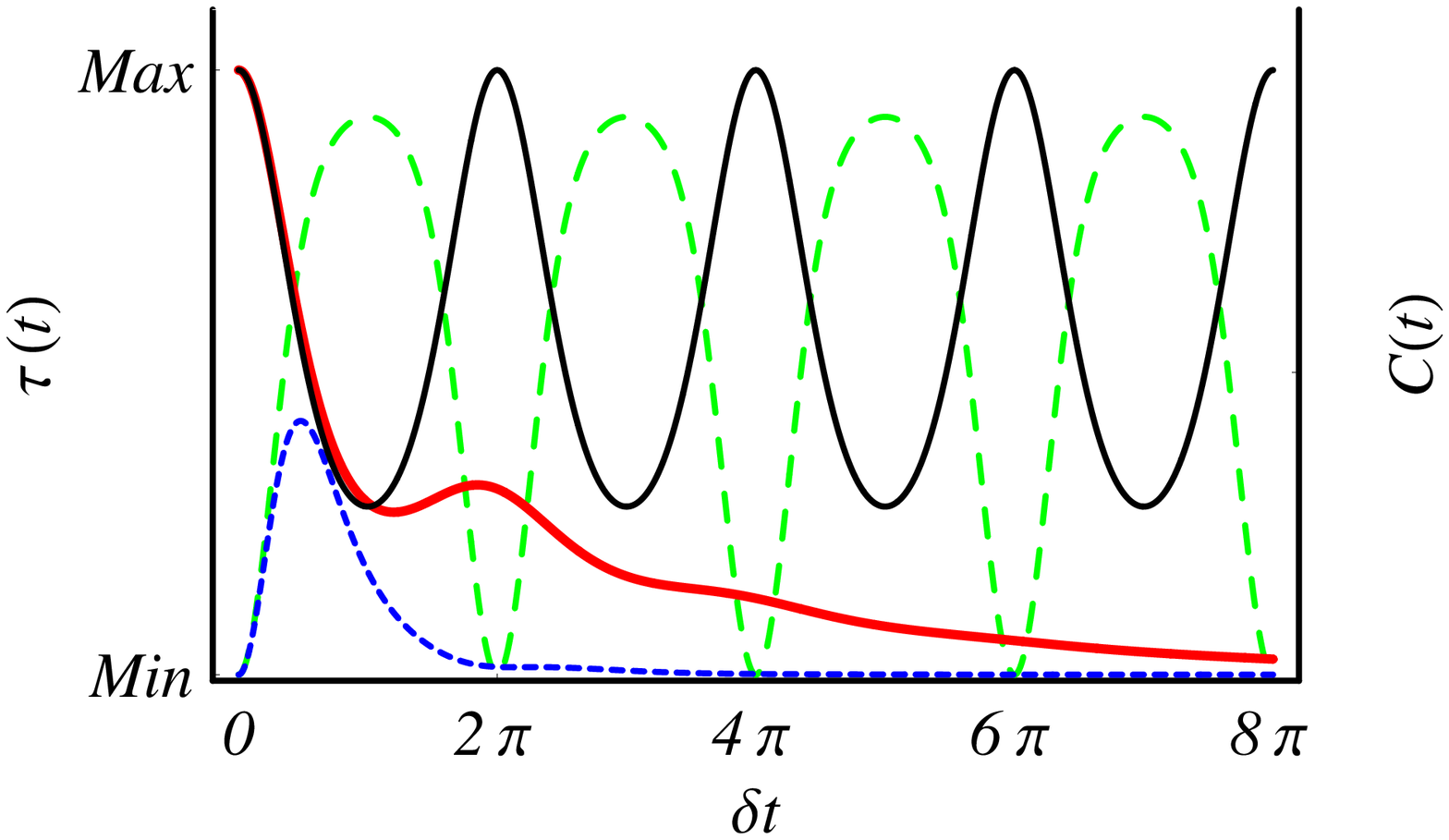}
	\caption{(Color online):  Tangles/qubit-qubit concurrences, without (green dashed/plum solid) and with (blue dashed/red solid) dissipation present made for $g_1=g_2=100$ MHz, $\delta=100$ MHz (first column),  $\delta=500$ MHz (second column), and $\k=10$ MHz (first row) and $\k=100$ MHz (second row) We see that the entanglement generated between the cavity and the two qubits depends on a delicate balance between the detuning and the dissipation rate. If detuning is small the tangle is also very short lived.}
	\label{fig:new3}
\end{figure*}

\section{Dissipative cavity} \label{sec:entevodiss}
Using the  solutions (\ref{solutions}), and repeating the analysis presented above in the dissipative cavity case we find  the negativities
\bea
{\rm Neg} \lr{Q_1 Q_2| C}&=&\frac{1}{4}\lr{e^{h_1}-1} \nnl &&\vspace{-1cm}+\frac{1}{4}\sqrt{1-4 e^{h_1}\lra{\chi }^2+2 e^{h_1}+e^{2 h_1}}\,,\nnl
{\rm Neg} \lr{Q_1 C| Q_2}\lr{t}&=&{\rm Neg} \lr{Q_2 C| Q_1}\lr{t}=\frac{1}{2}e^{h_1}\,,\nn
\eea
with $h_1=h_1\lr{g,t}$ is given by Eg. (\ref{h1}) and $\lambda$ stems from the definition (\ref{lambda}) and in this case reads
\bea
\lra{\chi }^2&=&\exp\lr{\frac{4 g^2 e^{-\k t} (\cos (\delta  t)-\cosh (\k t))}{\delta ^2+\k^2}}\nn \ .
\eea

We can see that as a result of cavity dissipation the negativity ${\rm Neg} \lr{Q_i C| Q_j}\lr{t}$ (constant when $\k=0$ becomes a nontrivial function of time whose plots are presented in Figure \ref{fig:new2} for different values of dissipation rate $\k$ and detuning $\delta$.
Other than the presence of detuning creating a (dis)entanglement oscillation, we see two competing effects playing a role here. Coupling-to-dissipation ratio at resonance leads to a decay of qubit-qubit entanglement and creation of qubits-cavity entanglement; detuning on the other hand, limits the qubit-qubit disentanglement, by means of impairing the qubit-cavity entanglement as we have seen in the previous section in Figure \ref{fig:gr1}. Small detunings, facilitate formation of coherent states of larger maximum amplitudes which make the state resemble the GHZ state to a larger extend, thus disentangling the qubits (lowering the qubit-qubit concurrence). When at the same time cavity dissipation rate is higher then the coherent state formed decreases its amplitude since $\a\propto\ g/(\k+i \delta)$ resulting in a slower monogamy of entanglement induced disentanglement rate.  For large detunings, the qubits are coupled to the cavity field less, causing a smaller coherent state amplitudes and reducing the disentanglement rate. Greater cavity decay rate again only amplify this process. This explains the findings of the previous paper \cite{Dukalski2010,Dukalski2011}, and we see this process quite clearly here due to detuning which marks the frequency of re- and disentanglement. Additionally, we point out that of the qubit-qubit (non)-dissipative concurrences
\bea
\mathcal{C} \lr{t}&=& C\lr{0} e^{-2\lr{g_1+g_2}^2t^2}\,,\nnl
\mathcal{C}_{\k}\lr{t}&=& C\lr{0} \exp\lrb{\frac{4\lr{g_1+g_2}^2}{\k^2}\lr{1-\k t-e^{-\k t}}}\,,\nn
\eea
it is the second one that decays slower in time. Thus the cavity dissipation process is not the cause for the qubit-qubit disentanglement, but rather it is the factor to decreases the initial disentanglement rate (Figure \ref{fig:new2}). We can see this clearly, since the concurrence function $\mathcal{C}_{\k}$, is always decreasing $\lr{\dot{\mathcal{C}}_{\k}\propto -g^2\k^{-1}}$ with a long time limit equal to zero for any finite positive value of $\k$, however for infinite $\k$ concurrence remains constant. This shows that in this system the large cavity decay prohibits the qubits from entangling to the cavity and protects the bipartite entanglement that the qubits share.

To get a fuller picture of the negativity time evolution, we would need to find the three-tangle $\tau_{\k}$ (where we use the $\k$ subscript to denote the dissipative case) which is
\bea
\tau_{\k}= e^{h_1\lr{t}}\sqrt{1-\lra{\chi }^2}\nonumber\,.
\eea

The conclusion that we can draw from these results is that  in either scenario there is entanglement being formed between the qubits and the cavity and as a consequence of monogamy of entanglement, the greater the degree of entanglement between the  qubits and the cavity the less can they retain their inner-qubit entanglement.  With the decay in the cavity present the entanglement between the qubits and the cavity weakens, which is a consequence of the cavity decay which can be seen as the cavity field entangling to the states of the environment which since traced over (a procedure carried out when deriving the Lindblad term), result in a continuous degradation of any entanglement present in the system as a \emph{whole}.

\section{Conclusions}\label{conclusion}

In this article we have studied the dynamics of tripartite entanglement between two driven qubits non-resonantly coupled to a  cavity. Using tripartite entanglement measures (negativity and three-tangle) we have shown that the previously reported entanglement loss followed by its revival is a consequence of an entanglement formation and subsequent disentanglement between the subsystem composed of qubits and the subsystem spanned by the cavity. Additionally, further  tripartite entanglement loss is due to the dissipation of the cavity, which can be seen to form a greater entangled state with the environment states which has been traced over in a process of a derivation the master equation. From this one can see the qubits coupled to the cavity, which acts as an intermediate non-Markovian bath, which is further coupled to a Markovian one. This non-Markovian-like behaviour can be seen to be due to the presence of the  external driving field and the cavity frequency mismatch $\delta$, which clocks the (dis)entanglement process. With this work we want to emphasize the danger of attributing \emph{all} correlation losses to dissipation \emph{alone} as seen by the evolution of correlations in this system, where the qubit-qubit entanglement is lost due to the qubits-cavity formation even for $\k=0$ and it was only the formation of a larger system-environment entanglement formation that lead to tripartite intra-system entanglement loss.

\section{Acknowledgements}
Authors would like to thank Antonio Borras Lopez, and Giorgi Labadze for useful discussions. this work was supported by the Netherlands Foundation for Fundamental Research on Matter (FOM).


\begin{thebibliography}{26}
\bibitem[{\citenamefont{comment1}}]{comment1}
\bibinfo{author}{\bibfnamefont{We assume that such an entangled state is available. In our follow up work we show how one could produce an entangled state using a system of  strongly driven qubits.}}
\end{thebibliography}
\end{document}